\documentclass[12pt,preprint]{aastex}
\usepackage{psfig}

\begin{document}

\setlength{\baselineskip}{12pt}

\newcommand\bb[1] {   \mbox{\boldmath{$#1$}}  }
\newcommand\del{\bb{\nabla}}
\newcommand\bcdot{\bb{\cdot}}
\newcommand\btimes{\bb{\times}}
\newcommand\vv{\bb{v}}
\newcommand\B{\bb{B}}
\newcommand\BV{Brunt-V\"ais\"al\"a\ }
\newcommand\iw{ i \omega }
\newcommand\kva{ \bb{k\cdot v_A}  }
\newcommand\beq{\begin{equation}}
\newcommand\eeq{\end{equation}}
\newcommand\QQ{ {\cal P} }

\def\dd{\partial}
\shortauthors{Balbus}
\shorttitle{Energy Transport in Nonradiative Accretion}
\slugcomment{Submitted to ApJ.}

\title{Turbulent Energy Transport\\
in Nonradiative Accretion Flows}

\author{Steven A. Balbus}

\affil{Dept. of Astronomy, University of Virginia, PO Box
3818,\\ Charlottesville VA 22903, USA. sb@virginia.edu}

\bigskip \begin{abstract} Just as correlations between fluctuating radial
and azimuthal velocities produce a coherent stress contributing to the
angular momentum transport in turbulent accretion disks, correlations in
the velocity and temperature fluctuations produce a coherent energy flux.
This nonadvective thermal energy flux is always of secondary importance
in thin radiative disks, but cannot be neglected in nonradiative flows,
in which it completes the mean field description of turbulence.  It is,
nevertheless, generally ignored in accretion flow theory, with the
exception of models explicitly driven by thermal convection, where it is
modeled phenomenologically.  This flux embodies both turbulent thermal
convection as well as wave transport, and its presence is essential for a
proper formulation of energy conservation, whether convection is present
or not.  The sign of the thermal flux is likely to be outward in real
systems, but the restrictive assumptions used in numerical simulations may
lead to inward thermal transport, in which case qualitatively new effects
may be exhibited.  We find, for example, that a static solution would
require inward, not outward, thermal transport.  Even if it were present,
thermal convection would be unlikely to stifle accretion, but would simply
add to the outward rotational energy flux that must already be present.

\end{abstract}


\section{Introduction}

\noindent The fate of gas in the vicinity of a black hole remains a
topic of intense astrophysical interest.   The classical motivation for
black hole accretion is of course the possibility of prodigious energy
output \citep{l69}, but in more recent years attention has also been
drawn to a class of flows that radiate inefficiently \citep{acls88,
ny95, nia00}.  With the unambiguous Chandra finding that the Galactic
Center source Sgr A$^*$ is severely underluminous \citep{bag01}, these
models have now acquired a compelling observational motivation.

Analytic models of nonradiating accretion flows (hereafter NRAFs) are
generally based on the classical $\alpha$ prescription, whereby all
transport is subsumed into a single enhanced viscosity parameter.
NRAFs are very likely to be turbulent, of course, because the
combination of differential rotation and a magnetic field is prone to the
magnetorotational instability, or MRI \citep{bh91}.  The simplifying
assumption that has almost universally been made is that the sole
dynamical effect of this turbulence is to act as the enlarged $\alpha$
viscosity parameter.  In earlier NRAF models, this assumption applied both
to energy as well as to angular momentum transport \citep{ny94, bb99}.
More recently, a class of accretion flow has appeared that specifically
invokes thermal convection as the dominant mode of energy transport,
while at the same time arguing that the angular momentum stress tensor
effectively vanishes \citep{nia00}.

The purpose of this paper is to elucidate, from first principles,
the nature of turbulent transport couplings in an NRAF.  The formalism
that allows a connection to be made between phenomenological transport
modeling and the fundamental fluid equations is weak turbulence theory.
This is a less restrictive limitation than it may sound.  It is equivalent
to retaining the quadratic correlations in the turbulent fluctuations, and
is a mathematically self-consistent approach.  Phenomenological models of
turbulent accretion flows certainly do not attempt higher order accuracy,
and often settle for far less.  

Applications of this type of formalism to nonradiative flows were
carried out by \cite{qn99}, and to black hole accretion more generally
by \cite{ketal98}.  But in its emphasis upon modes of transport rather
than difficult questions of closure, our approach more closely follows
that of \cite{bp99}.  These authors used weak turbulence theory to
elucidate the foundations of $\alpha$ modeling in thin radiative disks,
and found that whereas MHD processes naturally lend themselves to such a
description, self-gravity generally does not.  A related finding of this
study was that the {\em only} correlation tensor of dynamical importance
that emerges in an MHD turbulent Keplerian disk is the Maxwell-Reynolds
stress.  This fact, in turn, can be traced to the assumption that the
ratio of the isothermal sound speed $c_S$ to Keplerian rotation velocity
$v_K$ satisfies ${c_S/v_K}\ll 1$.  When this assumption no longer holds,
other transport coefficients become dynamically significant (as happens
in self-gravitating disks), in which case the simple viscous model itself
breaks down.

The regime ${c_S/v_K}\sim 1$ coincides with the NRAF regime.  We are
led, therefore, to consider flows in which the energy extracted from the
differential rotation is, in addition to being locally dissipated, {\em
nonlocally} transported by other correlations associated with a turbulent
or wavelike energy flux.  This form of transport has the potential to be
important in any NRAF, whether or not thermal convection is associated
with the accretion flow.  Its role is quite distinct from that of the
dynamical stress tensor, and the two should not be confused.

An outline of this work is as follows.  Section 2 presents a simple
discussion of correlated fluctuations and mean flow, in particular
emphasizing the circumstances in which fluctuations, not just mean flow
values, must be retained in the governing equations.  Section 3 focuses on
turbulent energy transport.  The key result is equation (\ref{fluctherm}),
which is the fundamental thermodynamic relation for the fluctuations.
Section 4 develops simple, one-dimensional (1D) solutions to the governing
equations, contrasting them with solutions from the literature.  Finally,
\S 5 is a summary of the paper.

\section {Correlated Fluctuations and Mean Flow}

\noindent

As stated in the Introduction, our presentation is based on weak
turbulence theory, which bridges the divide between phenomenological
and first-principle approaches to turbulent flow.  Any flow quantity
may be decomposed into the sum of a mean value plus a fluctuation with
zero-mean.  We use the notation  $\langle X\rangle$ for the mean value
of quantity $X$, and $\delta X$ to denote its fluctuating component.
For the flow velocity, weak turbulence means that the fluctuations are
small compared to the isothermal sound speed, $c_S$.  For the pressure
and density, the fluctuations are to be considered small compared with
the mean value.  The magnetic field $\bb{B}$ is assumed to be subthermal,
i.e., the Alfv\'en speed $v_A \sim \delta v \ll c_S$.  In what follows, we
shall not need to distinguish between the mean and fluctuating components
of $\bb{B}$.  Mean velocity components may be either very large or very
small compared with their RMS fluctuations.  The rotational velocity
will generally greatly exceed its azimuthal velocity fluctuation, but
the mean radial drift velocity, a consequence of dissipation, will be
of second order (i.e., quadratic) in the fluctuation amplitude.

This last point is key, and merits some discussion.
Upon long term time-averaging, the governing
equations of a turbulent flow 
have a leading asymptotic form which is either zeroth or
second order in the amplitudes.  It is important to distinguish these two
cases, because by definition
fluctuations play no role in the leading behavior of
a zeroth order equation, whereas fluctuations must be self-consistently
retained {\em everywhere} in a second
order equation.  Consider first the exact radial equation of
motion expressed in the usual $(R, \phi, Z)$ cylindrical coordinates:
\beq
\left( {\dd\  \over \dd t} + \bb{v \cdot \nabla}\right) v_R -
{v_\phi^2\over R} = -{1\over \rho} {\dd P \over \dd R} -{\dd\Phi\over \dd
R} +{1\over c} \bb{(J\times B)}_{\bb{R}}.
\eeq 
Our notation is standard: $\bb{v}$ is the velocity (subscripts denote
components), $\rho$ the mass density, $P$ the gas pressure, $\Phi$
the external gravitational potential, and $\bb{J}$ the current density.
If the mean radial velocity is of second order and the magnetic field
is weak, the time-steady leading order form of this equation is simply
\beq
{R \Omega^2 } = {1\over \langle\rho\rangle} {\dd\langle P
\rangle \over \dd R} +{\dd\Phi\over \dd R},
\eeq
where $\Omega$ is the mean angular velocity and all other terms are
leading zeroth order mean values.  In accord with familiar treatments of
accretion flows, fluctuations play no role here.
(A similar argument can be made for the $z$ equation of hydrostatic
equilibrium.)  Contrast this, however, with the
azimuthal equation of motion:
\beq \left( {\dd\  \over \dd t} + \bb{v \cdot \nabla}\right) v_\phi +
{v_R v_\phi\over R} = -{1\over \rho R} {\dd P \over \dd\phi} 
 +{1\over c} \bb{(J\times B)}_{\bb{\phi}}.  \eeq 
Here, there are no leading terms of zeroth order: after averaging, the
leading behavior of the equation is second order in the amplitudes (Balbus
\& Papaloizou 1999).  All terms involve either the second order radial
drift velocity $\langle v_R\rangle$, or explicit amplitude correlations
of the form $\langle \delta X\, \delta Y\rangle$.  These latter turbulent
correlations must be retained, or the resulting equation is incorrect.
Indeed, the correlated fluctuations in this equation
are just the (nonadvective) angular momentum flux, which is critical to
the accretion process itself.  Standard $\alpha$ treatments in effect
retain these correlations by inventing an ``anomalous viscosity.''
To the extent that the theory is independent of the functional form
of the correlations, the fact that at least something is present
in the equation where the turbulent stress should appear is sufficient
for some purposes.  It allows points of principle to be
illustrated (e.g., the initial spreading and subsequent concentration of
an evolving disk), as well as some robust results to be obtained
(e.g., the luminosity--accretion rate relation [Pringle 1981].)

Genuine
difficulties with the standard treatment emerge when one attempts to go
beyond rotational mechanics.  Consider the entropy equation:
\beq\label{s}
{P\over \gamma - 1}\, 
\left( {\dd\  \over \dd t} + \bb{v \cdot \nabla}\right)\ln P\rho^{-\gamma}
= Q^+ - Q^-,
\eeq
where $Q^+$ and $Q^-$ are respectively the volume specific entropy gains
and losses.  The tacit assumption that is almost always made is that
the flow quantities appearing in equation (\ref{s}) may be replaced by
their mean values.  This is incorrect.

The problem is that the mean radial velocity is a second order
quantity, and the {\em entire} entropy equation is therefore also of second
order.   The quantity $Pv_R$ is not, on average, simply the product of
the mean pressure and radial drift velocity, but (in one dimension):
\beq
\langle P v_R\rangle  = \langle \rho \theta v_R\rangle =
\langle \theta \rangle \langle \rho v_R \rangle + \langle\rho\rangle
\langle\delta v_R\, \delta \theta \rangle + {\rm higher\ order\ terms,}
\eeq
where $\theta\equiv c_S^2$ is the temperature variable.  In fact, the full
entropy equation is even more complex, because there are correlations
between $\delta v_R$ and the fluctuation of the entropy gradient term
itself that must be taken into account.  There is no {\em a priori} reason
that all of these correlation coefficients should vanish.  Clearly, care
must be taken when determining quantities such as 
the sign of the mean entropy gradient in
a turbulent accretion flow.   A more systematic approach is needed.

\section {Turbulent Transport in NRAFs}

\subsection{ Second Order Conservative Transport Equations}

\noindent To investigate the behavior of a model flow with a turbulent
thermal energy flux in more detail, begin with the time-steady form
of the angular momentum conservation equation,
\beq
\bb{\nabla\cdot} \left(\rho R  v_\phi \bb{v} - {R\over 4\pi} B_\phi \bb{B},
\right) = 0 
\eeq
and the 
energy conservation equation:
\beq
\bb{\nabla\cdot} \left[\rho\bb{v}\left({v^2\over 2} + \Phi +{\gamma\theta\over
\gamma -1}\right) +{1\over4\pi}\bb{B\times(v\times B)}\right] = -Q^-.
\eeq
We have neglected the contribution of the particle viscosity and
(potentially more seriously) the thermal conductivity.  Throughout the remainder
of this paper,
{\em all equations are understood to be azimuthally averaged.} 
Unsubscripted boldface vectors and vector operators should be regarded
as poloidal.

We next expand all quantities into mean plus fluctuating components.
The largest terms in the energy flux are second order in the
fluctuating amplitudes.  (Recall that we are assuming that the 
Alfv\'en speed is of the same order as a kinetic velocity
fluctuation.)  The mass flux is
\beq
\langle \rho \bb{v} \rangle = \langle\rho\rangle \langle\bb{v}\rangle +
\langle\delta\rho \, \bb{\delta v}\rangle, 
\eeq
and satisfies
\beq\label{mass}
\bb{\nabla\cdot}\langle \rho \bb{v} \rangle = 0 .
\eeq
If we now define the stress vector $\bb{W}$ to be
\beq
\bb{W} = \langle \delta v_\phi \, \bb{\delta v} - {{B_\phi}\bb{B}\over
4\pi\rho} \rangle , 
\eeq
then the angular momentum equation becomes
\beq\label{angmom}
\bb{\nabla\cdot}\left(
\langle \rho \bb{v} \rangle R^2\Omega + \langle\rho\rangle R \bb{W}
\right) = 0 .   
\eeq
The energy equation, through second order in weak turbulence theory, reads:
\beq\label{energy}
\bb{\nabla\cdot}\left[
\langle \rho \bb{v} \rangle
\left(
{R^2\Omega^2\over 2}  + \Phi  + {\gamma\langle\theta\rangle\over \gamma - 1 }
\right) +\langle\rho\rangle R \Omega\bb{W} +
{\gamma\over\gamma -1} \langle\rho\rangle\langle\delta\theta\, \bb{\delta
v}\rangle\right]
= - Q^-
\eeq
The first group of terms multiplying the mass flux
$\langle\rho\bb{v}\rangle$ corresponds to the Bernoulli constant in
polytropic spherical flow.  In general, of course, it need not be
constant.  The key component of the energy flux that distinguishes
transport in a nonradiative flow from a thin Keplerian disk involves
only one new correlation product: the thermal
energy flux proportional to $\langle \delta\theta
\delta\bb{v}\rangle$.  This term is generally ignored in $\alpha$
models of nonradiative accretion flows.   Instead, these models retain
a form of the energy flux appropriate for a radiative thin disk, in
which the dominant nonadvective energy flux comes from the rotational
transport term $\bb{W}$.  In general, the ratio of the thermal
energy flux to rotational stress will
be, as noted, of order $c_S/R\Omega$.   While this justifies neglect of
the thermal energy flux in thin disk models, it also shows that this 
correlation must be retained in any flow with comparable rotation and
sound speed.  Indeed, retention of the energy flux terms is essential to
formulating a thermodynamically self-consistent model for the energetics
of turbulent fluctuations.

\subsection {Fluctuation Thermodynamics}

\noindent
The result of combining mass conservation (\ref{mass}) with radial 
hydrostatic equilibrium (2) in the 
energy conservation equation is
\beq
{\langle\rho v_R\rangle\over 2R^2}{d(R^4\Omega^2)\over dR} +
{\langle \theta\rangle \over \gamma -1}
\langle\rho \bb{v}\rangle  
\bb{\cdot \nabla} S
+ \bb{\nabla\cdot}\left(
\langle \rho \rangle  R \Omega \bb{W} + {\gamma\langle \rho \rangle\over \gamma -1}\langle\bb{\delta
v}\, \delta\theta\rangle \right)= -Q^-,
\eeq
where 
\beq
S \equiv \ln \left[ \langle P\rangle \langle\rho\rangle^{-\gamma}\right].
\eeq
If we now combine
this equation with angular momentum conservation (\ref{angmom}),
our result simplifies to
\beq \label {fluctherm} \bb{\nabla\cdot}\left(
{\gamma\langle\rho\rangle\over\gamma -1}\langle\bb{\delta v}\,
\delta\theta \rangle\right) + {\langle\theta\rangle\over\gamma -1}\langle
\rho\bb{v} \rangle \bb{\cdot\nabla} S = -\left(Q^-
+\langle\rho\rangle W_{R\phi} {d\Omega\over d\ln R}\right) \eeq

Equation (\ref{fluctherm}), which is the internal energy equations for the
fluctuations, is a key theoretical result of this paper.  It is readily
interpreted.  The right hand side is the rate at which fluctuations
exchange energy with their ``surroundings.''  The first term ($Q^-$)
represents bulk radiative losses, and the final term is the rate at which
energy is supplied by the free energy reservoir of differential rotation.
In a classical thin Keplerian disk, these terms together comprise the
dominant energy balance, and the entire left side of the equation may
be ignored.

In a nonradiative flow, however, none of the free energy of differential
rotation escapes, but is instead routed through two possible
thermodynamic paths.  The first possibility, embodied by the divergence
term on the left side of (\ref{fluctherm}), is direct mechanical
transport of the thermal energy content of a fluctuation, including
simple advection plus $PdV$ work.  The form of the flux is
$$
{\gamma\langle\rho\rangle\over \gamma -1} \langle\delta\theta\, \bb{\delta v}
\rangle =  \langle\rho\rangle C_P \langle \delta T\,\bb{\delta v} \rangle,
$$
where $C_P$ is the mass specific heat capacity at constant pressure
(cf.\ Schwarzschild 1958).  In the case of an adiabatic sound wave, this
expression is equivalent to $ \langle\delta P\, \bb{\delta v}\rangle$,
precisely the formal rate of work done by the wave.  The balance set
by equating this term with the final term on the right side of equation
(\ref{fluctherm}) corresponds to wave action conservation for adiabatic
disturbances \citep{l78}.

The second term on the left is the average rate at which entropy is
generated by the mean mass flow.  It is the only other thermodynamic
channel available to the free energy of differential rotation:
if the energy is not carried off by adiabatic processes, then it
must be dissipated.  In turn, the dissipated energy must, in the
absence of radiative losses, drive either an inflow or outflow.
Dynamical considerations strongly favor an outflow, at least in the
outer flow regions, leading to the scenario envisaged by Blandford \&
Begelman (1999).  To the extent that bulk flow does {\em not} occur,
a thermal flux must carry the energy away.  This term does not appear
in the Blandford \& Begelman formulation, so this form of transport is
precluded in their analysis.

Finally, it should be noted that equation (\ref{fluctherm}) differs
somewhat from the internal equation (4) used in \cite{aiqn02} and
elsewhere by the same authors.  This is because the fluctuation formalism
is not used by \cite{aiqn02}, so the internal energy input is regarded as
a source of particle heating, and is therefore dissipative.  By contrast,
in this work we have granted the fluid the degrees of freedom associated
with fluctuations, which allows the free energy of differential rotation
to be extracted with or without irreversible dissipation.

\subsection{Thermal Convection vs.\ Thermal Flux}

The discussion of the previous section invites comparison with the
convection-dominated flows put forth \cite{nia00}.  While the dynamical
description of these models has been criticized, (Balbus \& Hawley 2002;
Narayan et al.\ 2002 for a reply), one should not dismiss the possibility
that a thermal energy flux might affect the flow structure under more
general circumstances.  The point goes beyond the presence or absence
of thermal convection, and allows us to sharpen the very notion of what
it means for a flow to be ``convection-dominated.''

In an accretion flow whose primary source is the free energy of
differential rotation, what does it mean to say that the turbulence is
dominated by thermal convection?  The unstable linear mode in question
is {\em always} the same slow mode branch of the MHD dispersion relation
regardless of the mixture of adverse shear and thermal gradients that may
be present.  (Recent claims to the contrary in the literature reflect
confusion on this point.)  If the primary instability is the MRI,
which in its nonlinear resolution allows accretion to proceed, and any
adverse entropy gradients are present only as a secondary consequence of
accretion, in what sense can the flow ever be dominated by convection?

As originally formulated \citep{nia00, aiqn02}, the role of convection
was truly dominant: it was argued that it could cause the stress tensor
to vanish and that it could eliminate turbulent dissipation throughout
the body of the accretion flow.  But this leads to serious difficulties
(Balbus \& Hawley 2002).  The vanishing of the radial component of
$\bb{W}$ (more precisely a density-weighted average thereof, denoted
$T_{R\phi}$) leaves the fluctuations without a local source of free
energy.  Source-free, dissipation-free local turbulence maintained
by large dissipative entropy gradients is a thermodynamically
dubious proposition.  It is therefore not surprising that global,
three-dimensional MHD accretion simulations \citep{hbs01, hb02, ina03}
all report robust positive values of $T_{R\phi}$.

More recently, claims that a simulated accretion flow is dominated by
convection have shifted to morphological grounds \citep{ina03}, with only
the visual appearance of the flow used as a basis for an identification.
This is not the diagnostic of choice.  Whether or not thermal convection
is of significance in a high temperature NRAF simulation is in fact a
well-posed quantitative problem.  One must first begin by treating energy
and angular momentum transport on the same footing by elevating the energy
correlation flux {$\langle\delta\theta \bf{\delta v}\rangle$} to the
same status as $T_{R\phi}$.  Second, controlled studies are essential.
The visual appearance of turbulence is not a very reliable guide to the
presence or absence of thermal convection, especially if the root cause
of the turbulence is a magneto-rotational process.  Simulations must
be run with and without a dissipative heat source, and both turbulent
correlation coefficients examined in detail for each case.  This has
in fact already been done for $T_{R\phi}$: \cite{sp01} and \cite {hb02}
both found very little difference in the behavior of the stress tensor
whether or not resistive heating was present in their simulations.
The turbulent energy flux has yet to undergo a similar level of scrutiny,
however.  The question is, how much does this thermal flux correlation
change relative to its value in the absence of a dissipative heat source?
What matters ultimately is whether the thermal energy flux is an efficient
drain for the free energy of differential rotation in any nonradiative
flow, convective or otherwise.

\section {One-Dimensional Solutions}

\subsection {Formulation of the Problem}

\noindent 
To better understand the physical consequences of a thermal energy flux
in an NRAF, we apply the formal results of \S 3 to some explicit, simple
accretion solutions.  To do this, we need explicit functional forms for
$\bb{W}$ and $\langle\delta\theta \bb{\delta v}\rangle$.  Since our goal
is one of illustrating a point of principle, we will use the dimensional
analysis associated with $\alpha$ formalism for each of these fluxes.
That is,
\beq
\langle \rho \delta v_R\, \delta v_\phi -
{B_R\, B_\phi\over4\pi} \rangle = \alpha_{SS}\langle P \rangle,
\quad \langle \rho \delta v_R\, \delta \theta \rangle = \alpha_{T}
\langle P\rangle \langle c_S\rangle
\eeq
where $\alpha_{SS}$ (the Shakura-Sunyaev $\alpha$ parameter) and
$\alpha_T$ are dimensionless constants.  Note that $\alpha_{SS} >0$,
but we leave the sign of $\alpha_T$ unconstrained for the moment.
The dynamical stress must be positive in order to extract energy from
the differential rotation, whereas the the thermal flux is a secondary
consequence of the turbulence, and could in principle have either sign,
depending on what mode of transport dominates.  This is particularly
important when one uses this flux as a numerical diagnostic,
because the restrictions sometimes used (e.g. two dimensions, adiabatic
gas law) can bias the direction of the energy flux.

Next, we take the rather drastic but standard step of height-integrating
the equations of motion, and assume that what remains is a well-defined,
1D (on average) radial flow \citep{ny94, bb99}.  It is useful to
have such a formal solution to compare with others in the literature.
Numerical NRAF simulations do show a well-defined thick wedge forming
along the midplane with a relatively autonomous interior, surrounded
by a low density coronal envelope \citep{hb02}.  There is some sense
in regarding much of the wedge structure as 1D, although there is also
a significant amount of high latitude backflow that is lost in this
approximation.  Perhaps the best reason to study these 1D flows is to
illustrate a point of principle when they fail to work self-consistently,
in contrast, say, to steady viscous disk models.

Our 1D formulation is similar to the ``idealized ADAF''
analyzed by \cite{bb99}, with an added thermal energy flux.  The equations
of motion, which can now be expressed in terms of mean value quantities,
are
\beq\label{1dr}
\Sigma R\Omega^2 = \int {d\langle P\rangle\over dR}\, dZ + {GM\Sigma\over R^2}
\eeq
\beq\label{1da}
- { {\dot m} \Omega \over 2\pi} + \alpha_{SS} \QQ = {C_J\over R^2}
\eeq
\beq\label{1de}
-{ {\dot m}\over 2\pi}\left( {R^2\Omega^2\over 2} -{GM\over R}
+ {\gamma \over \gamma - 1 } {{\cal P}\over \Sigma} \right) +
R^2\Omega\alpha_{SS} \QQ + {R\gamma\over \gamma -1}\alpha_T {\QQ^{3/2}
\over \Sigma^{1/2}} = C_E
\eeq
These are, respectively, the radial equation of motion, and the angular
momentum and energy conservation equations.  ${\dot m}$ is the conserved mass
accretion rate, $C_J$ is the conserved rate of angular momentum transport, and
$C_E$ is the conserved energy transport rate.  $\Sigma$ is the height
integrated density (i.e., the column density), and $\QQ$ is the height
integrated pressure.  We have assumed a central black hole mass $M$, but
as is customary, have used a gravitational potential depending only upon $R$.
An alternative useful form for equation (\ref{1de}) 
is obtained by eliminating the $\alpha_{SS}$ term via equation
(\ref{1da}):
\beq\label{1debis}
{ {\dot m}\over 2\pi}\left( {R^2\Omega^2\over 2} +{GM\over R}
- {\gamma \over \gamma - 1 } {{\cal P}\over \Sigma} \right) +
{R\gamma\over \gamma -1}\alpha_T {\QQ^{3/2}
\over \Sigma^{1/2}} = C_E - \Omega C_J
\eeq

\subsection{Static Envelope}

\noindent We begin with a 
search for solutions with ${\dot m} = 0$.  
Combining equations (\ref{1dr}) with (\ref{1da}) gives
\beq\label{disc1}
\QQ= {C_J\over \alpha_{SS} R^2}, \quad \langle P\rangle \propto R^{-3},
\eeq
and
\beq
\Omega^2 = {GM\over R^3} - {3C_J\over \alpha_{SS}\Sigma R^4}. 
\eeq
The radial power law dependence of the pressure suggests we look
for similar behavior in $\Omega$ and $\Sigma$.  This is possible
only if $C_E =0$, so that the net energy flux vanishes.  
Then equation (\ref{1debis}) leads directly to
\beq\label{omR}
\Omega = - {\gamma\over \gamma - 1 }{\alpha_T\over \alpha_{SS}^{3/2}}
{C_J^{1/2}\over \Sigma^{1/2}   R^2}.
\eeq
Evidently, it is necessary that $\alpha_T <0$, i.e. the heat transport must be
{\it inward} to ensure there is no energy deposition.  
Eliminating $\Omega$ between the last two equations leads to an equation
for $\Sigma$:
\beq\label{Sigma}
\Sigma = {3C_J\over \alpha_{SS}GMR}( 1 + \xi), \quad \xi \equiv
{\alpha_T^2\gamma^2 \over  3\alpha_{SS}^2(\gamma-1)^2 },
\eeq
and $\Omega$ follows from (\ref{omR}), 
\beq\label{Omega}
\Omega^2 = {\xi \over 1+\xi}{GM\over R^3}.
\eeq
This is a sub-Keplerian profile and is associated with thick disk structure
(Frank, King, \& Raine 2002).
The temperature is given by
\beq\label{theta}
\langle \theta \rangle = {\QQ\over \Sigma} = {GM\over 3R}( 1 + \xi)^{-1}
\eeq
Equations (\ref{Sigma}--\ref{theta}) comprise our 1D
static solution.  Notice that it satisfies the virial equilibrium condition
\beq
2T + \Phi + 3\langle \theta \rangle = 0
\eeq
where $T=R^2\Omega^2/2$ is the rotational kinetic energy, and that it differs
from the static convection dominated solutions of \cite{nia00} and 
\cite {aiqn02} in requiring an inward, rather than an outward, thermal flux.
The cause of this difference is that our solution has a positive
value of the $R\phi$ stress tensor component, as is required by energy
conservation.

\subsection{Accreting Envelope}

\noindent Equations (\ref{1dr}--\ref{1de}) allow for power law solutions
even when ${\dot m}$ does not vanish, a property shared with NRAFs in
their standard formulation \citep{ny94, bb99}.  We require the vanishing
of both the energy as well as the angular momentum flux, $C_E=C_J =0$.
Since $W_{R\phi}$ is no longer directly proportional to $C_J$, as in
the static solution of the previous section, there is no difficulty with
simultaneously demanding a positive stress and a vanishing net angular
momentum flux.   This condition is satisfied in classical Keplerian
disks at radii large compared with the inner edge.

With 
\beq
\QQ \propto R^{-3/2}, \quad \Sigma \propto R^{-1/2}, \quad \Omega
\propto R^{-3/2},
\eeq
equation (\ref{1dr}) becomes
\beq
\langle \theta \rangle
= {\QQ\over \Sigma} = {2\over 5}\left( {GM\over R} - R^2\Omega^2
\right)
\eeq
and after some manipulations, the energy equation (\ref{1de}) may
be written
\beq
\label{ref1}
{R^2\Omega^2\over 2}\left( {9\gamma\over 5} - 1 \right)
- {GM\over R} \left( 1 -{3\gamma\over 5}\right) +
{\gamma \alpha_T\over \alpha_{SS}} 
\left(
2 R^2 \Omega^2\over 5\right)^{1/2} \left(
{GM\over R} - R^2\Omega^2 \right)^{1/2} = 0.
\eeq
Defining
\beq
A= {9\gamma\over5} - 1, \quad B=1 - {3\gamma\over5}, \quad C =
\gamma{\alpha_T\over\alpha_{SS}}\left(2\over5\right)^{1/2},
\quad x^2 = {R^3\Omega^2\over GM},
\eeq
we find that $x$ must satisfy
\beq
x^4(C^2 + A^2/4) - x^2(C^2 +AB) +B^2 = 0.
\eeq
The choice of sign for this quadratic (in $x^2$) equation 
must be chosen to be consistent with equation (\ref{ref1}),
a spurious root having been introduced in the course of squaring a radical.

It is the appearance of $C$ that distinguishes 
our solution from those of earlier NRAF studies.  If $C>0$, corresponding to
outward transport, then it must
be large compared with $A$ to change
the $C=0$
solution $x^2=2B/A$ qualitatively.  (The novel $B$ = 0 solution
of equation [\ref{ref1}] is spurious when 
$C>0$.)  On the other hand, if $C<0$, corresponding to inward
transport, a
qualitative change does occur: whereas a standard 1D
NRAF has no solution for the important case $\gamma = 5/3$
(i.e. $B=0$), the presence of an inward energy flux leads
to a new sub-Keplerian solution
\beq
x^2 = (1+A^2/4C^2)^{-1}, \quad \theta =
(2/5)(GM/R)(1+4C^2/A^2)^{-1}.
\eeq
An {\em inward} thermal flux seems anomalous.  Considering also its
novel consequences for both static and accreting flows, we are thus led
to consider the physical processes leading to either inward or outward
transport.

\subsection{Discussion}

\noindent 
It is perhaps surprising that, in these 1D models, {\em inward} thermal
transport is required to establish either a static profile, or a $\gamma
=5/3$ accretion profile.  But given a power law leading asymptotic order
behavior for the thermal and rotational energy fluxes at large $R$, there
is little choice.  Energy must be strictly conserved.  Therefore, the
two fluxes must either cancel one other, or their sum must be conserved.
Because it forces a unique power-law behavior, the latter possibility
overconstrains the problem.  Hence the rotational energy flux, which
must always be outward, is canceled by an equal and opposite thermal flux.

Second, equation (\ref{fluctherm}), together with static or isentropic
flow, implies a very frugal use of free energy: essentially all of it
must go into an inward thermal flux.  What might be the physical basis
for such behavior?

The MRI, acting in the presence of a Schwarzschild-stable entropy
gradient, will tend to drive a thermal energy flux {\em down} the
entropy gradient, which in this case is indeed inwards \citep{b00}.
For the static solution, equations (\ref{disc1}) and (\ref{Sigma}) give
$$ S \sim \ln R^{2\gamma -3} $$ 
for the static solution, which
is Schwarzschild-stable for $\gamma > 1.5$.  On the other hand, for the
$\gamma =5/3$ accreting solution, $S$ is constant with $R$.

But there are deeper problems with the mechanism of turbulent mixing,
even for the $\gamma
> 1.5$ static solution.  The inward energy transport is produced by an
intrinsically dissipative process, and the entropy of a fluid element
would continuously increase.   Without radiative losses, a static
solution would not persist under these circumstances.  This is reflected
in equation (\ref{fluctherm}), where it may be seen that the dissipative
term is coupled both to mass flow as well as to a finite entropy gradient.

Might it be possible to continuously extract the free energy of
differential rotation and have it go directly into {\em nondissipative}
energy transport?  Certainly.  This is what waves do in the process of
conserving their action, and non-dissipative wave propagation does not
cause systematic mass flow in the background medium.  But in our problem,
any waves that are present will be coupled to dissipative turbulence.

This raises another issue.  In general, the linear behavior of a
rotating magnetized fluid is marked by two incompressible modes.  At a
fixed wavelength and arbitrarily weak field strengths, one branch is
a destabilized slow mode (the MRI); the other is generally a stable
inertial wave (assuming that entropy gradient is dominated by its
vertical component), which becomes Alfv\'enic as the field strengthens.
Nonlinear coupling between the unstable and stable modes is almost
certain to be present in a turbulent fluid.

Hydrodynamical inertial waves in disks are characterized by frequencies
$\omega$ below the epicyclic value \citep{vd89, b03}, which in a Keplerian
disk is simply $\Omega$.  Hence axisymmetric waves propagate inwards,
because any initially outward propagating wave at fixed $\omega$ will
eventually encounter a turning point at which $\omega$ exceeds $\kappa$,
and the wave must reflect.  Two-dimensional numerical simulations of the
MRI would then be characterized by $\alpha_{SS} > 0$ (outward transport
by rotational stress) and, to the extent that inertial forces exceed
those of magnetic tension, $\alpha_T < 0$.  The MRI dominates the angular
momentum transport (axisymmetric inertial waves carry no angular momentum
of course), but perhaps not the energy transport.  A preliminary numerical
investigation of a global, two-dimensional, self-gravitating, adiabatic,
magnetized torus does indeed show comparable values of $\alpha_{SS}$ and
$|\alpha_T|$, with $\alpha_T < 0$ (Fromang, Devilliers, \& Balbus 2003).

In
three dimensions, the situation is significantly more complex, because inertial
waves are not free to propagate inwards without reflection.
Rather than $\omega < \kappa$, the condition for propagation is
$$
|\omega - m \Omega| < \kappa,
$$
where $m$ is the azimuthal wave number.  The radius at which equality
in the above relation is obtained is known as the Lindblad resonance,
and normally there are two such locations: outer and inner Lindblad
resonances, where $\omega - m \Omega$ is respectively positive and
negative.  The existence of an inner Lindblad resonance is a consequence
of nonaxisymmetry,
and the resulting trapping of any inertial waves undermines
the arguments used in the axisymmetric case.  One may always appeal to the
magnetic field to get passed the Lindblad barriers, but there is no guarantee
that waves liberated in this way will be heading preferentially inwards.

The case for inward thermal energy transport is even more problematic
in real black hole systems.  Coulomb conduction, heretofore ignored, is
likely to be an important process in the dilute plasmas that characterize
nonradiative black hole accretion.  The relevant instability criterion for
a plasma whose thermal conductivity is dominated by a magnetized Coulomb
conductivity is that the {\em temperature}, not the entropy, increase
outward \citep{b01}.  The combined magnetorotational/magnetothermal
instability would result in outward, not inward, thermal transport.
A static solution under these conditions is not possible.

Both acoustic waves and turbulence associated with an unstable thermal
gradient will tend to move thermal energy outward with high efficiency,
since in both cases the correlation between the velocity and temperature
fluctuation is strong.  In the case of acoustic waves, a net outward flux
is expected since initially inward propagating waves will tend to refract
and head outwards.  Outwardly propagating waves, on the other hand, will
become increasingly dominated by their radial wavenumber, and eventually
dissipate or shock.  This would seem more likely to result in a wind from
the outer regions of the flow, rather than a static or accreting halo.

To conclude, in a height-integrated 1D analytic formalism,
inward transport of thermal energy is needed for a static solution or
for $\gamma =5/3$ rotating accretion to exist.   Conditions favorable to
inward transport include axisymmetric and adiabatic flow, restrictive
assumptions likely to be more common in numerical simulations than in
nature.  Three-dimensional flow, and the mixing of magnetothermal with
magnetorotational instabilities in the presence of a Coulomb conductivity,
all tend to produce an outward energy flux.  In the presence of a positive
rotational stress, outward thermal transport does not stifle accretion.
It does, however, lead to accretion flows whose salient properties
cannot be captured by 1D height-integrated modeling.  

\section{Summary}

\noindent The key point of this paper is that thermal fluctuations
must be taken into account in the formulation of governing equations of
nonradiative accretion flows, and in particular that
the entropy equation cannot be analyzed
directly by assuming all flow quantities take on their mean values.  The energy
flux 
\beq\label{keypt} {\gamma\over\gamma - 1}\langle  \rho \rangle
\langle \delta \theta\, \bb{\delta v} \rangle \eeq
is an important
component in any nonradiative accretion process.  Energy conservation,
the central assumption of nonradiative flows, is not satisfied unless this
term is present, and such basic processes as wave action conservation
are lost.  It is important here to note that the presence of this
correlation flux is more fundamental than the mere inclusion of a some
additional transport process, such as thermal conduction.  The issue is
one of thermodynamic self-consistency.  The energy flux (\ref{keypt}) {\em
must} accompany the advected flux term
\beq {\gamma\langle \theta\rangle
\over\gamma - 1} \langle \rho \bb{v} \rangle \eeq
for the same reasons, and at the same level of formalism, that the
stress tensor $\bb{W}$ accompanies the advected angular momentum flux.
The results are summarized quantitatively in equation (\ref{fluctherm}),
which expresses energy conservation for either wavelike or turbulent
fluctuations.

With one exception, no other accretion model familiar to the author
includes a nonadvective thermal energy flux.  As noted in \S 3.1,
convection-dominated models \citep{nia00, qg00, aiqn02} assume that
the energetics are dominated by an outward energy flux generated by
a convectively unstable accretion process.  The central assumption of
these models is that convection inhibits angular momentum transport and
causes the turbulent stress to vanish.  We have argued, on the other
hand, that the stress tensor $W_{R\phi}$ must be positive in order to
extract free energy from differential rotation.  One consequence of
this is that any viable, height-integrated static solution must have an
inward flux of thermal energy, rather than an outward flux, as expected
in a convectively unstable flow.   (The same inward flux requirement
holds for the existence of steady, height-integrated, $\gamma=5/3$
rotating accretion flows.)  It therefore seems unlikely that thermal
convection, even if it were present, would stifle accretion.  Instead,
it would simply add to the outward rotational energy transport while
negligibly affecting the angular momentum transport \citep{sp01, hb02}.

Establishing how the thermal flux emerges in a mean flow formalism for
weak turbulence is a theoretical result that we believe will foster a
deeper understanding of nonradiative accretion, if for no other reason
then for avoiding the pitfall of confusing flow quantities with their
mean values.  But the immediate utility of the present work is likely
to be as a numerical diagnostic.  This is particularly true for those
cases in which thermal convection is claimed to be present.  
to the Reynolds/Maxwell stress tensor, the energy correlation $\langle
\bb{\delta v}\, \delta\theta \rangle$ has not been well-studied.  It is
particularly important to know how much of the free energy is ultimately
``radiated'' by this under-explored form of mechanical luminosity.

\acknowledgments{It is a pleasure to thank J.~Hawley and an anonymous
referee for detailed and constructive comments on an earlier draft
of this paper.  I am also grateful to G. Dubus, S. Fromang, and J.-P.
Lasota for stimulating conversations and suggestions, and to the Institut
d'Astrophysique de Paris, whose hospitality and support allowed much
of this work to be completed.  This work was supported by NASA grants
NAG5--9266 and NAG5--13288, and NAG--10655.

\end{document}